\documentclass[aps,prc,floatfix,onecolumn,nofootinbib,superscriptaddress,notitlepage]{revtex4-1}

\usepackage[]{graphicx}% Include figure files
\usepackage{xcolor}

\usepackage{amsmath,amssymb,amsfonts}

\usepackage{slashed}

\usepackage{color}
\usepackage[utf8]{inputenc}
\allowdisplaybreaks
\usepackage[bookmarks,
                bookmarksopen = true,
                bookmarksnumbered = true,
                linktocpage,
                colorlinks = true,
                linkcolor = blue,
                urlcolor  = blue,
                citecolor = blue,
                anchorcolor = green,
                hyperindex = true,
                hyperfigures]
                {hyperref}

\begin{document}

\title{Photon bremsstrahlung from heavy quarks in a dense nuclear matter}

\author{Le Zhang}
\email{zlphys@csuc.edu.cn}
\affiliation{School of General Education, Cyberspace Security University of China, Wuhan, 430040, China}
\affiliation{College of Physics and Electronic Science, Hubei Normal University, Huangshi, 435002, China}

%\affiliation{Institute of Particle Physics and Key Laboratory of Quark and Lepton Physics (MOE), Central China Normal University, Wuhan, 430079, China}

\author{Shanshan Cao}
\email{shanshan.cao@seu.edu.cn}
\affiliation{School of Physics, Southeast University, Nanjing 211189, China}

\author{De-Fu Hou}
\email{houdf@mail.ccnu.edu.cn}
\affiliation{Institute of Particle Physics and Key Laboratory of Quark and Lepton Physics (MOE), Central China Normal University, Wuhan, 430079, China}

\author{Guang-You Qin}
\email{guangyou.qin@ccnu.edu.cn}
\affiliation{Institute of Particle Physics and Key Laboratory of Quark and Lepton Physics (MOE), Central China Normal University, Wuhan, 430079, China}

\date{\today}
%%%%%%%%%%%%%%%%%%%%%%%%%%%%%%%%%%%%%%%%%%%%%%%%%%%%%%%%%%%%%%%%%%%%%%%%%%%%%%%%
%%%%%%%%%%%%%%%%%%%%%%%%%%%%%%%%%%%%%%%%
\begin{abstract}

We study the bremsstrahlung photon production from a hard jet parton induced by rescattering with a dense nuclear medium. Using the charged current interaction channel of deep inelastic scattering between an electron and a large nucleus, we derive the spectrum of medium-induced photons emitted from high-energy heavy and light quarks at the next-to-leading twist in a unified framework. Going beyond the collinear expansion approximation, we show that the photon spectrum is determined by the full momentum distribution of the gluon exchanged between the propagating quark and the medium, or equivalently, by the differential elastic scattering rate of the hard quark inside the medium. Modeling the gluon field with a static Debye screened potential reduces the photon spectrum to a dependence on the transverse momentum distribution of the exchanged gluon. This work provides a more reliable input for future phenomenological studies of quark mass effects on jet-induced photon production in relativistic heavy-ion collisions. 

\end{abstract}
\maketitle
%%%%%%%%%%%%%%%%%%%%%%%%%%%%%%%%%%%%%%%%%%%%%%%%%%%%%%%%%%%%%%%%%%%%%%%%%%%%%%%%
%%%%%%%%%%%%%%%%%%%%%%%%%%%%%%%%%%%%%%%%

\section{Introduction}

Jets produced from the early stage high-energy (hard) partonic scatterings between nuclei are regarded as a powerful tool to probe the properties of the hot and dense nuclear matter created in high-energy nuclear collisions, as conducted at the Relativistic Heavy-Ion Collider (RHIC) and the Large Hadron Collider (LHC)~\cite{Wang:1991xy,Qin:2015srf, Blaizot:2015lma,Cao:2020wlm,Wang:2025lct,Mehtar-Tani:2025rty}.
During their propagation through the dense nuclear medium, jet partons interact with the medium via elastic scatterings and their induced gluon emissions (inelastic scatterings), and therefore lose energy and forward momentum before exiting the medium and fragmenting into hadrons. A direct consequence of these jet-medium interactions is the yield suppression of high transverse momentum ($p_\mathrm{T}$) hadrons and jets in relativistic heavy-ion collisions compared to that in proton-proton collisions scaled by the number of binary nucleon-nucleon collisions~\cite{Adcox:2001jp,Adler:2002xw,Aamodt:2010jd,CMS:2012aa,Abelev:2012hxa,Aad:2015wga}, a phenomenon called jet quenching.

From the theoretical side, tremendous efforts have been devoted to understanding the parton energy loss and jet modification in a dense (either cold or hot) nuclear matter. For example, the collisional energy loss experienced by the hard partons traversing the dense nuclear medium has been studied in Refs.~\cite{Bjorken:1982tu,Braaten:1991we,Djordjevic:2006tw,Qin:2007rn}. Meanwhile, the medium-induced radiative energy loss of hard partons has been investigated in various formalisms, such as Baier-Dokshitzer-Mueller-Peigne-Schiff-Zakharov (BDMPS-Z)~\cite{Baier:1996kr,Baier:1996sk,Baier:1998kq,Zakharov:1996fv,Zakharov:1997uu}, Djordjevic-Gyulassy-Levai-Vitev (DGLV)~\cite{Gyulassy:1999zd, Gyulassy:2000er, Djordjevic:2003zk},
Armesto-Salgado-Wiedemann (ASW)~\cite{Wiedemann:2000za,Wiedemann:2000tf,Armesto:2003jh}, Arnold-Moore-Yaffe (AMY)~\cite{Arnold:2001ba,Arnold:2002ja} and higher twist (HT)~\cite{Guo:2000nz,Wang:2001ifa,Zhang:2003wk,Majumder:2009ge,Du:2018yuf}.
A detailed comparison between these different approaches can be found in Ref.~\cite{Armesto:2011ht} and references therein. Based on these theoretical formalisms, sophisticated numerical studies have been performed to investigate the effects of jet-medium interactions on various final state observables, such as the yield suppression and anisotropic flow coefficients of hadrons and jets, di-jet, di-hadron, and hadron-jet correlations, and intra-structures of jets~\cite{Cao:2024pxc,Cao:2022odi}. One of the main purposes of these studies is to utilizing jets to quantitatively extract thermal and transport properties of the QGP, including the jet transport coefficient~\cite{Baier:1996sk,JET:2013cls,JETSCAPE:2021ehl,Xie:2022ght,Xie:2024xbn} and heavy quark diffusion coefficient inside the QGP, and the viscosity~\cite{Karmakar:2023ity} and the equation of state~\cite{Feal:2019xfl,Liu:2023rfi} of the QGP. 

Current calculations on the radiative energy loss of jet partons mainly focus on gluon emissions induced by the transverse momentum broadening of jet partons as they propagate through the dense nuclear medium. When a jet parton scatters off the medium constituents, both its transverse and longitudinal momenta are altered via gluon exchange with the medium~\cite{Majumder:2008zg,Qin:2012fua,Abir:2014sxa}. Effects of the longitudinal momentum loss experienced by the jet parton have been investigated in several phenomenological studies, but within the context of evaluating purely collisional energy loss of either leading hard partons~\cite{Wicks:2005gt,Djordjevic:2006tw,Qin:2007rn,Schenke:2009ik} or radiated gluons within full jet showers~\cite{Qin:2009uh,Neufeld:2009ep,Qin:2010mn,Qin:2012gp,Chang:2016gjp}.
Recently, effects of the longitudinal momentum exchange on the photon and gluon emission processes have been investigated in Refs.~\cite{Qin:2014mya,Abir:2015hta,Zhang:2016avg,Zhang:2018kkn,Zhang:2018nie,Sirimanna:2021sqx,Kumar:2025egh,Kumar:2025asj} by allowing the longitudinal momentum exchange to be on the order of the transverse momentum exchange. In this work, we study the real photon bremsstrahlung from a heavy or light quark jet scattering off a dense nuclear medium with both transverse and longitudinal momentum transfers. This is an extension of the series of our previous studies~\cite{Zhang:2016avg,Zhang:2018kkn,Zhang:2018nie}. In Ref.~\cite{Zhang:2016avg}, we applied the collinear expansion approximation (expansion about the transverse momentum exchange) and derived the medium-induced single photon bremsstrahlung spectrum from a light flavor jet, including the contributions of the longitudinal drag, longitudinal diffusion, and transverse broadening. In Refs.~\cite{Zhang:2018kkn,Zhang:2018nie}, we worked beyond the collinear expansion approximation and derived a closed formula for the medium-induced gluon emission spectrum from a light or heavy quark jet interacting with a dense nuclear medium via both transverse and longitudinal rescatterings. 
Compared to Ref.~\cite{Zhang:2016avg}, we attempt to derive the medium-induced single photon bremsstrahlung spectrum from a heavy or light quark jet on the same footing beyond the collinear expansion approximation, with both transverse and longitudinal momentum transfers included. We find that the medium-induced photon spectrum depends on the full momentum distribution of the exchanged gluon, which is equivalent to the differential scattering rate of the jet parton inside the nuclear medium.
The single photon emission spectrum obtained in this work can be used as an input to evaluate the production of jet-induced photons in relativistic heavy-ion collisions, which are expected to provide a significant contribution to the direct photon spectrum in the intermediate $p_\mathrm{T}$ regime~\cite{Fries:2002kt,Qin:2009bk}.
Furthermore, as a special case, we utilize the static Debye screened potential to model the exchanged gluon field and reduce our results to a medium-induced single photon bremsstrahlung spectrum that only depends on a distribution function of the transverse momentum of the exchanged gluon. Our conclusion is consistent with the description of the gluon bremsstrahlung process in Refs.~\cite{Zhang:2018nie,Zhang:2019toi}.

The remainder of this paper is organized as follows.
In Sec.~\ref{sec:II}, we first briefly discuss the photon bremsstrahlung process from a heavy quark jet within the framework of semi-inclusive deep inelastic scattering in vacuum, which contributes to the hadronic tensor at the leading twist level. 
In Sec.~\ref{sec:III}, we investigate the medium-induced photon bremsstrahlung resulting from a single scattering of the heavy quark inside a dense nuclear medium (at the next-to-leading twist level). Working beyond the collinear expansion limit, we provide the relation between the single photon bremsstrahlung spectrum and the full momentum distribution of the exchanged gluon between the heavy quark and the medium. 
Sec.~\ref{sec:IV} contains our summary. Some details of our calculations are presented in the Appendix.

\section{Photon bremsstrahlung in vacuum}
\label{sec:II}

In this section, we first briefly discuss the photon bremsstrahlung process from a heavy or light quark jet in vacuum, produced within the framework of semi-inclusive deep inelastic scattering (DIS) off a large nucleus,
\begin{eqnarray}
e^-(L_1) + A(A p) \to \nu_e(L_2) + q(l_q) + \gamma(l) + X,
\end{eqnarray}
where $L_1$ and $L_2$ denote the momenta of the incoming and outgoing leptons, $l_q$ and $l$ are the momenta of the produced hard quark and the radiated photon, and $A\,p$ is the momentum of the incoming nucleus with an atomic number $A$. Each nucleon in the nucleus has a four-momentum $p = [p^+, m_N/(2 p^+), \mathbf{0}_\perp] \approx [p^+, 0, \mathbf{0}_\perp]$ with the light-cone notation, where $m_N$ is the mass of a nucleon which can be neglected in the high energy limit.
In this work, we consider the charged current interaction channel, which allows us to study medium-induced photon bremsstrahlung from heavy and light quarks on the same footing. In the charged current interaction channel, the exchanged $W^-$ boson carries a four-momentum $q = L_2 - L_1 = [-x_B p^+, q^-, \mathbf{0}_\perp]$, where $x_B = Q^2 / (2p^+q^-)$ is the Bjorken fraction variable with $Q^2=-q^2$ as the invariant mass of the exchanged $W^-$ boson, with $Q^2\ll m_W^2$ assumed. As an example, this $W^-$ boson can strike a $u$ quark inside the nucleus and convert it into a high-energy (hard) $d$, $s$, or $b$ quark. With this setup, the momentum of the incoming nucleus is dominated by its ``$+$" component, while the momentum of the hard quark (or jet) is dominated by its ``$-$" component (inherited from the $W^-$ boson).

In the semi-inclusive DIS process above, the differential cross section can be expressed as~\cite{Guo:2000nz, Zhang:2003wk}
\begin{eqnarray}
E_{L_2}  \frac{ d\sigma_\mathrm{DIS}}{d^3\mathbf{L}_2} = \frac{G^2_F}{(4\pi)^3 s}  L_{\mu\nu}   W^{\mu\nu},
\end{eqnarray}
where $G_F$ is the four fermion coupling constant and $s=(p+L_1)^2$ is the center-of-mass energy of the lepton-nucleon collision system. The charged current leptonic tensor $L_{\mu\nu}$ reads
\begin{eqnarray}
L_{\mu\nu} = \frac{1} {2} \rm Tr[\slashed{L}_1  (1+\gamma^5) \gamma_\mu \slashed{L}_2 \gamma_\nu (1-\gamma^5)],
\end{eqnarray}
and the hadronic tensor $W^{\mu\nu}$ is given by
\begin{equation}
W^{\mu\nu} =\frac{1}{2}\sum_{X} (2\pi)^4 \delta^4 (q + A p - P_X-l_q-l) \langle A| J^\mu(0)|X\rangle \langle X|{J^\nu}^\dag(0)|A\rangle,
\end{equation}
where $|A\rangle$ represents the initial state of an incoming nucleus $A$, $|X\rangle$ denotes the final hadronic (or partonic) states, and $\sum_{X}$ sums over all possible final states except the outgoing hard quark and the emitted photon.
The hadronic charged current reads $J^\mu = \bar{\psi}_i \gamma^\mu  (1-\gamma^5) V_{ij} \psi_j $, in which $V_{i j}$ is the Cabibbo-Kobayashi-Maskawa (CKM) matrix for flavor mixing~\cite{Aivazis:1993kh}.
In this work, our main focus is on the hadronic tensor $W^{\mu\nu}$, which encodes the final-state interaction between the propagating hard quark and the nuclear medium it traverses.

\begin{figure}[thb]
\includegraphics[width=0.6\linewidth]{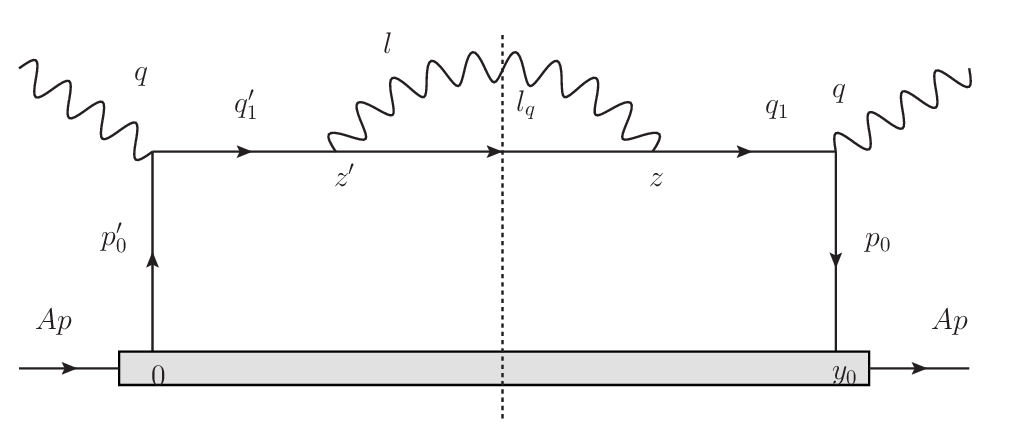}
\caption{ Photon bremsstrahlung process in vacuum.
} \label{fig1}
\end{figure}

In Fig. \ref{fig1}, we illustrate the photon bremsstrahlung process in vacuum within the framework of semi-inclusive DIS. The exchanged $W$ boson with momentum $q$ strikes a massless quark with momentum $p_0'$ ($p_0$ in the complex conjugate) from one nucleon of the nucleus at the location $y_0' = 0$ ($y_0$ in the complex conjugate) and produces a hard (either heavy or light) quark with mass $M$ and momentum $q_1'$ ($q_1$ in the complex conjugate). The hard quark emits a real photon at the location $z'$ ($z$ in the complex conjugate). The hard quark and its emitted photon carry momenta $l_q$ and $l$, respectively, both exiting the nuclear medium without further interaction with the medium. This process corresponds to the leading twist hadronic tensor, which reads
%\begin{widetext}
\begin{eqnarray}
\label{hadronictensorfig0}
W_0^{A \mu\nu} \!\!&=&\!\! \sum_q Q_q^2 e^2 \int \frac{d^4l}{(2\pi)^4} (2\pi)\delta(l^2) \int \frac{d^4l_q}{(2\pi)^4} (2\pi) \delta(l_q^2-M^2)
\int d^4y_0 e^{iq\cdot y_0} \int d^4z \int d^4z' \nonumber\\\!\!&\times&\!\!
\int \frac{d^4 q_1}{(2\pi)^4} \int \frac{d^4 q_1'}{(2\pi)^4} e^{-i q_1 \cdot (y_0-z)} e^{-i q_1' \cdot (z'-y_0')} e^{-il_q\cdot (z-z')} e^{-i l\cdot (z-z')} \nonumber\\\!\!&\times&\!\!
|V_{ij}|^2\langle A |\bar{\psi}_i(y_0) \gamma^\mu (1-\gamma^5) \frac{\slashed{q}_1+M}{q_1^2 -M^2- i\epsilon} \gamma^\alpha (\slashed{l}_q +M)\gamma^\beta\frac{\slashed{q}_1'+M}{q_1'^2 -M^2+ i\epsilon}(1+\gamma^5)\gamma^\nu \psi_i(0)|A\rangle G_{\alpha\beta}(l).
\end{eqnarray}
Here, the superscript $A$ of $W_0^{A\mu\nu}$ denotes the nucleus is composed of $A$ nucleons, $\mu$, $\nu$, $\alpha$, and $\beta$ are the Lorentz indices associated with the $W$ boson and photon fields at the locations of $y_0$, $y_0'$, $z$, and $z'$, respectively. We use the light-cone gauge, $n\cdot A = A^-=0$, with $n=[1,0,\mathbf{0}_\perp]$ the direction along which the high-energy nucleus propagates. The sum of the photon polarizations in the light-cone gauge takes the form of 
\begin{eqnarray}
G_{\alpha\beta}(l) = -g_{\alpha \beta} + \frac{n_\alpha l_\beta + n_\beta l_\alpha}{n\cdot l}.
\end{eqnarray}
In the limit of very high energy, one can neglect the $\perp$-components of the quark field operators and factor out the one-nucleon state from the nucleus as follows:
\begin{eqnarray}
\label{factor out aa}
& &\langle A | \bar{\psi}_i(y_0) \gamma^\mu (1-\gamma^5) \hat{O}   (1+\gamma^5)\gamma^\nu \psi_i(0) | A \rangle
\nonumber\\
&\approx&
A C_p^A \langle p | \bar{\psi}_i(y_0^-) \frac{\gamma^+}{2} \psi_i(0) | p \rangle\times\frac{1}{4 p^+ q^-} {\rm Tr} [\slashed{p} \gamma^\mu (1-\gamma^5) \{\slashed{q}+ (x_B + x_M)\slashed{p}\} (1+\gamma^5)\gamma^\nu  ]\times {\rm Tr} [\frac{\gamma^-}{2}  \hat{O} ],
\end{eqnarray}
where $C_p^A$ denotes the probability of finding a nucleon state with momentum $p$ inside the nucleus $A$, and $x_M = M^2/(2 p^+ q^-)$ is defined for convenience.

To simplify Eq.~(\ref{hadronictensorfig0}), one may reorganize its phase factors as $e^{- i (q_1-q) y_0} e^{-i (l+l_q -q_1)z}e^{-i (q_1'-l-l_q)z'}$. This yields two four-$\delta$-functions enforcing momentum conservation---$\delta(l+l_q-q_1)$ and $\delta(q_1'-l-l_q)$---after the photon radiation locations $z$ and $z'$ are integrated over.
Then, we may perform integrations over the four-momenta $q_1$ and $q_1'$, and set $q_1 = q_1' = l + l_q$. Furthermore, we re-introduce the variable $p_0$ from the following identity
\begin{eqnarray}
\int d^4 p_0 \delta(q +p_0 - l -l_q) = 1,
\end{eqnarray}
and use it to integrate out $l_q$ by setting $l_q=q+p_0-l$. Integrals over $y_0^+$ and $\mathbf{y}_{0\perp}$ yield a three-$\delta$-function, which can be used to integrate over $p_0^-$ and $\mathbf{p}_{0\perp}$. Consequently, we can set $p_0=[x_0 p^+, 0, \mathbf{0}_\perp]$ with $x_0 = p_0^+/p^+$. These simplifications reduce the hadronic tensor to
\begin{eqnarray}
\label{eq:WAsimplify1}
{W_{0}^{A \mu\nu} } &=& \sum_q Q_q^2 e^2 \int \frac{d^4l}{(2\pi)^4} (2\pi)\delta(l^2)
\int d y_0^- \int \frac{p^+  d x_0}{2\pi}e^{- ix_0 p^+ y_0^-}
\frac{1}{q_1^2 - M^2 - i\epsilon}\frac{1}{q_1^2 - M^2 + i\epsilon}
 (2\pi) \delta(l_q^2-M^2)
 \nonumber\\&\times&
|V_{ij}|^2 A C_p^A \langle p |\bar{\psi}_i(y_0^-) \frac{\gamma^+}{2} \psi_i(0)|p \rangle \times\frac{1}{4 p^+ q^-}{\rm Tr} [\slashed{p} \gamma^\mu (1-\gamma^5)  \{\slashed{q}+ (x_B + x_M)\slashed{p}\} (1+\gamma^5) \gamma^\nu ]
\nonumber\\&\times&{\rm Tr}[\frac{\gamma^-}{2}(\slashed{q}_1 + M)\gamma^\alpha  (\slashed{l}_q + M)
\gamma^\beta (\slashed{q}_1 + M)]{G}_{\alpha\beta} (l).
\end{eqnarray}

Now we can use the photon on-shell condition $\delta(l^2)$ to integrate over $l^+$, yielding $l^+ = l_\perp^2 / (2\,y q^-)$, where $y=l^-/q^-$ is the fractional forward momentum carried by the radiated photon from its parent quark. Similarly, we can integrate over the momentum fraction $x_0$ by using the on-shell condition of the final-state quark as
\begin{eqnarray}
\delta(l_q^2-M^2)=\frac{1}{2 p^+ q^-(1-y) }\delta(x_0 - x_B - x_M -\tilde{x}_L ),
\end{eqnarray}
in which the momentum fraction $\tilde{x}_L = (l_\perp^2 + y^2 M^2)/[2p^+q^-y(1-y)]$ can be related
to the formation time of the radiated photon from a heavy quark jet as $\tilde{\tau}_{\rm{form}}^-=1/(\tilde{x}_L p^+)=2q^-y(1-y)/(l_\perp^2+yM^2)$.

The trace part at the end of Eq.~(\ref{eq:WAsimplify1}) is evaluated as
\begin{eqnarray}
{\rm Tr} [ \frac{\gamma^-}{2} (\slashed{q}_1+M) \gamma^\alpha (\slashed{l}_q+M) \gamma^\beta (\slashed{q}_1+M) ] G_{\alpha\beta}(l)
=4 q^- P(y) \frac{l_\perp^2 +\frac{y^4}{1+(1-y)^2}M^2}{y (1-y)},
\end{eqnarray}
with $P(y)=[1+(1-y)^2]/y$ the quark-to-photon splitting function.

One may further define the light quark distribution function in a nucleon of the incoming nucleus as
\begin{eqnarray}
\label{eq:defPDF}
f_i(x) = \int \frac{d y_0^-}{2\pi} e^{-i x p^+y_0^-} \langle p| \bar{\psi}_i(y_0^-) \frac{\gamma^+}{2} \psi_i(0)|p \rangle,
\end{eqnarray}
in which $x$ denotes the fractional forward momentum carried by the light quark from the nucleon. This leads to the final expression of the differential hadronic tensor within the framework of semi-inclusive DIS at the leading twist as
\begin{eqnarray}
\label{eq:vaccum}
\frac{dW_0^{A\mu\nu}}{ dy dl_\perp^2 } &=& \sum_q Q_q^2 A C_p^A (2\pi) f_i(x_B + x_M + \tilde{x}_L)
\times\frac{1}{4 p^+ q^-} |V_{ij}|^2{\rm Tr} [\slashed{p} \gamma^\mu (1-\gamma^5) \{\slashed{q}+ (x_B + x_M)\slashed{p}\} (1+\gamma^5) \gamma^\nu]\nonumber\\
&\times&
\frac{\alpha_e}{2 \pi} {P(y)}\times  \frac{l_\perp^2 + \frac{y^4}{1+(1-y)^2}{M^2}}{(l_\perp^2  + y^2 {M^2})^2}.
\end{eqnarray}

By assuming $\tilde{x}_L\ll x_B$, one may extract the differential photon emission spectrum from a hard quark in vacuum as
\begin{eqnarray}
\frac{d N_\gamma^{\rm{vac}}}{d y d l_\perp^2 } = \frac{\alpha_e}{2 \pi} {P(y)} \frac{l_\perp^2 + \frac{y^4}{1+(1-y)^2}{M^2}}{(l_\perp^2  + y^2 {M^2})^2}.
\end{eqnarray}
With the quark mass $M$ set to zero, the expression above is reduced to
\begin{eqnarray}
\frac{d N_\gamma^{\rm{vac}}}{d y d l_\perp^2 } = \frac{\alpha_e}{2 \pi} \frac{1}{l_\perp^2} {P(y)} ,
\end{eqnarray}
same as the single photon emission spectrum from a light quark in vacuum obtained in Refs.~\cite{Qin:2014mya, Zhang:2016avg}. In the next section, we will investigate the effect of a single scattering experienced by a massive hard quark in a dense nuclear matter on this photon bremsstrahlung spectrum.
%\end{widetext}

\section{Photon bremsstrahlung in a dense nuclear matter}
\label{sec:III}

\begin{figure}[thb]
\centering
\includegraphics[width=0.6\linewidth]{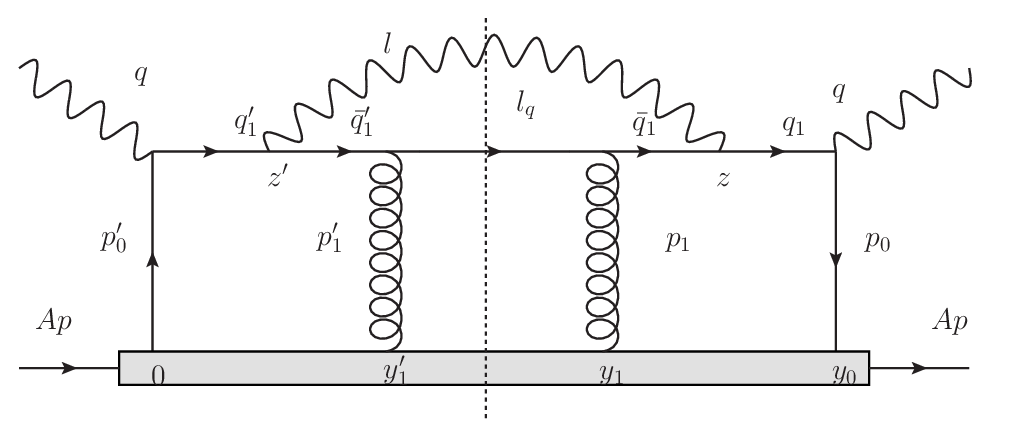}
\caption{A central-cut diagram: one rescattering on the hard quark after photon emission in both the amplitude and its complex conjugate.} \label{fig2}
\end{figure}

In this section, we derive the single photon radiation spectrum induced by a single scattering in a dense nuclear medium. This corresponds to the twist-four contribution to the hadronic tensor compared to the leading twist process presented in the previous section.
Here, we present one of the central-cut diagrams at the twist-four level in Fig.~\ref{fig2} and compute its hadronic tensor in detail. There are 3 other central-cut diagrams and 6 non-central-cut diagrams, whose evaluations are completely analogous and will be shown in the Appendix.

In Fig.~\ref{fig2}, a $W$ boson strikes a quark with momentum $p_0'$ ($p_0$ in the complex conjugate) from the nucleus at location $y_0' = 0$ ($y_0$ in the complex conjugate) and converts it into a hard quark (jet parton) with mass $M$ and momentum $q_1'$ ($q_1$ in the complex conjugate). This hard quark propagates through the nuclear medium and emits a real photon with momentum $l$ at location $z'$ ($z$ in the complex conjugate). After the photon emission, the momentum of the hard quark changes to $\bar{q}_1'$ ($\bar{q}_1$ in the complex conjugate). It further scatters with the nuclear medium by exchanging momentum $p_1'$ ($p_1$ in the complex conjugate) at location $y_1'$ ($y_1$ in the complex conjugate) and ends with an outgoing (on-shell) momentum $l_q$.
The hadronic tensor for Fig.~\ref{fig2} can be written as
%\begin{widetext}
%
\begin{eqnarray}
\label{hadronictensorfig1}
W^{A\mu\nu}_{(\ref{fig2})} \!\!&=&\!\! \sum_q Q_q^2 e^2 g^2 \frac{1}{N_c} {\rm Tr}\left[T^{a_1} T^{a_1'} \right]
\int \frac{d^4l}{(2\pi)^4} (2\pi)\delta(l^2) \int \frac{d^4l_q}{(2\pi)^4} (2\pi)\delta(l_q^2-M^2) \int d^4y_0 e^{iq\cdot y_0} \int d^4y_1\int d^4y_1'
\nonumber\\\!\!&\times&\!\!
\int d^4z \int d^4z'\int \frac{d^4q_1}{(2\pi)^4} e^{-iq_1\cdot (y_0-z)} \int \frac{d^4\bar{q}_1}{(2\pi)^4} e^{-i\bar{q}_1\cdot (z-y_1)}e^{-il\cdot (z-z')} e^{-il_q\cdot (y_1-y_1')}\int \frac{d^4\bar{q}_1'}{(2\pi)^4} e^{-i\bar{q}_1'\cdot (y_1'-z')}
\nonumber \\\!\!&\times&\!\!
\int \frac{d^4q_1'}{(2\pi)^4} e^{-iq_1'\cdot z'}|V_{ij}|^2 \langle A | \bar{\psi}_i(y_0) \gamma^\mu(1-\gamma^5) \frac{\slashed{q}_1 + M}{q_1^2 - M^2 - i\epsilon}\gamma^\alpha \frac{\slashed{\bar{q}}_1 + M}{\bar{q}_1^2 - M^2 - i\epsilon} \slashed{A}_{a_1}(y_1)
\nonumber\\\!\!&\times&\!\!
(\slashed{l}_q + M)
\slashed{A}_{a'_1}(y'_1) \frac{\slashed{\bar{q}}_1' + M}{\bar{q}_1'^{2} - M^2 + i \epsilon}
\gamma^\beta \frac{ \slashed{q}_1' + M}{{q}_{1}'^2 - M^2 + i \epsilon}
(1+\gamma^5) \gamma^\nu \psi_i(0) |A\rangle
G_{\alpha\beta}(l),
\end{eqnarray}
in which $\mu$, $\nu$, $\alpha$, and $\beta$ are the Lorentz indices associated with the $W$ boson and photon fields at $y_0$, $y_0'$, $z$, and $z'$, respectively, $a_1$ and $a_1'$ are the color indices of the gluon fields at $y_1$ and $y_1'$, respectively. The subscript of $W^{A\mu\nu}_{(\ref{fig2})}$ denotes that this hadronic tensor corresponds to the process illustrated in Fig.~\ref{fig2}.

Similar to the evaluation of Eq.~(\ref{hadronictensorfig0}) in the previous section, we first reorganize the phase factors and integrate over the locations $z$ and $z'$, leading to two four-$\delta$-functions---$\delta(-q_1+\bar{q}_1+l)$ and $\delta(-l-\bar{q}'_1+q'_1)$---which are further used to integrate over the momenta ${q}_1$ and ${q}_1'$, leaving $q_1=\bar{q}_1+l$ and $ q_1'=\bar{q}_1'+l$.
Then, we re-introduce the momentum variable $p_0$ with %{\color{gray}($p_0 = q + p_1 - l - l_q$)} 
\begin{equation}
\int d^4 p_0 \delta(p_0+q+p_1-l_q-l)=1,
\end{equation}
and use the $\delta$-function to integrate out $l_q$.
Using the momentum conservation at each interaction vertex, one can write down the following relations for various momenta in Fig. \ref{fig2}:
\begin{eqnarray}
p_0=q_1-q, \,\,\,\,\,\, p_1=l_q-\bar{q}_1, \,\,\,\,\,\,
p'_0=q'_1-q, \,\,\,\,\,\, p'_1=l_q-\bar{q}'_1.
\end{eqnarray}
By changing the integration variables from $\bar{q}_1$ to $p_1$ and from $\bar{q}'_1$ to $p'_1$, the phase factors of Eq.~(\ref{hadronictensorfig1}) are reduced to $e^{-ip_0\cdot y_0} e^{-ip_1\cdot y_1} e^{ip'_1\cdot y'_1}$.

%In this work, we perform the calculation in the light-cone gauge ($A^-=0$) in the limit of very high energy and collinear radiation. Therefore, the dominant component of the vector potential is the forward $(+)$-component, which means that only $(-)$-component of $\gamma$ matrices contribute.
In the high-energy limit, we ignore the $\perp$-components of the quark and gluon field operators inside the nucleus, and factor out the one-nucleon state from the nucleus as follows:
\begin{eqnarray}
& &\langle A | \bar{\psi}_i(y_0) \gamma^\mu(1-\gamma^5) {(\slashed{q}_1 + M)}\gamma^\alpha {(\slashed{\bar{q}}_1 + M)} \slashed{A}_{a_1}(y_1)
(\slashed{l}_q + M)\slashed{A}_{a'_1}(y'_1) {(\slashed{\bar{q}}_1' + M)}
\gamma^\beta { (\slashed{q}_1' + M)}(1+\gamma^5) \gamma^\nu \psi_i(0) |A\rangle
\nonumber\\
&\approx& A C_p^A \langle p | \bar{\psi}_i(y_0^-) \frac{\gamma^+}{2} \psi_i(0) | p \rangle \times\frac{1}{4 p^+ q^-} {\rm Tr} [\slashed{p} \gamma^\mu (1-\gamma^5) \{\slashed{q}+ (x_B + x_M)\slashed{p}\} (1+\gamma^5) \gamma^\nu]
\nonumber\\
& & \times
\langle A | A_{a_1}^+(y_1) A_{a'_1}^+(y'_1) |A\rangle {\rm Tr} [\frac{\gamma^-}{2} (\slashed{q}_1+M) \gamma^\alpha (\slashed{\bar{q}}_1+M) \gamma^- (\slashed{l}_q+M) \gamma^- {(\slashed{\bar{q}}_1'+M)}
\gamma^\beta (\slashed{q}_1'+M)],
\end{eqnarray}
in which the light-cone gauge ($A^-=0$) is applied. With this, Eq.~(\ref{hadronictensorfig1}) is reduced to
\begin{eqnarray}
W^{A\mu\nu}_{(\ref{fig2})} \!\!&=&\!\! \sum_q Q_q^2 e^2 g^2 \frac{1}{N_c} {\rm Tr}\left[T^{a_1} T^{a_1'} \right] \frac{1}{4 p^+ q^-}|V_{ij}|^2 {\rm Tr} [\slashed{p} \gamma^\mu (1-\gamma^5) \{\slashed{q}+ (x_B + x_M)\slashed{p}\} (1+\gamma^5) \gamma^\nu]
\nonumber\\ \!\!&\times&\!\!
\int \frac{d^4l}{(2\pi)^4} (2\pi)\delta(l^2)\int d^3 \mathbf{y}_0 \int \frac{d^3\mathbf{p}_0}{(2\pi)^3} \int d^3 \mathbf{y}_1\int \frac{d^3\mathbf{p}_1}{(2\pi)^3}\int d^3 \mathbf{y}_1'\int \frac{d^3\mathbf{p}_1'}{(2\pi)^3}e^{-i\mathbf{p}_0\cdot \mathbf{y}_0} e^{-i\mathbf{p}_1\cdot \mathbf{y}_1} e^{i\mathbf{p}_1'\cdot \mathbf{y}_1'}
\nonumber\\ \!\!&\times&\!\!
\int d y_0^- \int d y_1^- \int d y_1'^- A C_p^A \langle p | \bar{\psi}_i(y_0^-) \frac{\gamma^+}{2} \psi_i(0) | p \rangle \langle A | A_{a_1}^+(y_1) A_{a'_1}^+(y'_1) |A\rangle
\nonumber\\ \!\!&\times&\!\!
\int \frac{d p_0^+}{2 \pi} \int \frac{d p_1^+}{2 \pi} \int \frac{d p_1'^+}{2 \pi}(2\pi)\delta(l_q^2-M^2)
\frac{e^{- i x_0 p^+ y_0^-}}{q_1^2-M^2-i\epsilon}\frac{e^{- i x_1 p^+ y_1^-}}{\bar{q}_1^2-M^2-i\epsilon}
\frac{e^{i x_1' p^+ y_1'^-}}{\bar{q}_1'^2-M^2+i\epsilon}\frac{1}{q_1'^2-M^2+i\epsilon}
\nonumber\\ \!\!&\times&\!\!
{\rm Tr} [\frac{\gamma^-}{2} (\slashed{q}_1+M) \gamma^\alpha (\slashed{\bar{q}}_1+M) \gamma^- (\slashed{l}_q+M) \gamma^- {(\slashed{\bar{q}}_1'+M)}
\gamma^\beta (\slashed{q}_1'+M)]
G_{\alpha\beta}(l).
\end{eqnarray}
For convenience, we use the three-vector notations for the momentum and coordinate space variables as $\mathbf{p} = (p^-, \mathbf{p}_\perp)$ and $\mathbf{y} = (y^+, \mathbf{y}_\perp)$, with their dot product given by $\mathbf{p}\cdot \mathbf{y} = p^-y^+ - \mathbf{p}_\perp \cdot \mathbf{y}_\perp$. The notations $x_0=p_0^+/p^+$, $x_1=p_1^+/p^+$, and $x_1'=p_1'^+/p^+$ are also used. One can integrate first over $\mathbf{y}_0$ and then over $\mathbf{p}_0$, rendering the constraint $\mathbf{p}_0=0$ for the high-energy limit of the incoming nucleus.

%{\color{blue}In the limit of very high energy and collinear radiation, we may further carry out the integrations over the location $\mathbf{y}_0$ and the momentum $\mathbf{p}_0$, rendering the momentum $\mathbf{p}_0=0$.}

We further simplify the denominators of the quark propagators and the on-shell condition of the outgoing quark as follows,
\begin{eqnarray}
& &q_1^2 - M^2 = (q + p_0)^2 -M^2 =  2 p^+ q^- [-{x}_B + x_0 - x_{M}],
\nonumber\\
& &\bar{q}_1^2-M^2 = (q+p_0-l)^2-M^2 = 2p^+q^-(1-y)[-x_B + x_0 -{x}_{M}- \tilde{x}_L ],
\nonumber\\
& &l_q^2-M^2 = (q+p_0+p_1-l)^2-M^2= 2p^+q^-(1+ x_1^- -y)[-x_B + x_0+{x}_1 + y\,x_M - (1-y)\tilde{x}_L- x_{C1}],
\end{eqnarray}
in which we introduce the momentum fractions
\begin{eqnarray}
 {x}_0 = \frac{p_0^+}{p^+} , \quad {x}_1 = \frac{p_1^+}{p^+}, \quad {x}_1^- = \frac{p_1^-}{q^-}, \quad x_{C1} = \frac{(\mathbf{l}_{\perp} - \mathbf{p}_{1 \perp})^2+M^2}{2p^+q^-(1 + x_1^- -y )}.
\end{eqnarray}
The momenta $q_1'$ and $\bar{q}_1'$ can be treated similarly.
Combining the contributions from the denominators of all the internal quark propagators and the on-shell condition for the outgoing quark, we obtain: 
\begin{eqnarray}
D_q &=& C_q (2\pi) \delta(-x_B + x_0+{x}_1 + y\, x_M - (1-y)\tilde{x}_L- x_{C1})
\nonumber\\&\times&
\frac{1}{-{x}_B + x_0 - x_{M}-i\epsilon} \frac{1}{-x_B + x_0 - {x}_{M}- \tilde{x}_L -i\epsilon}
\frac{1}{-{x}_B + x_0' - x_{M} +i\epsilon} \frac{1}{-x_B + x_0' -{x}_{M}- \tilde{x}_L+i\epsilon},
\end{eqnarray}
with
\begin{eqnarray}
C_q=\frac{1}{(2p^+q^-)^5(1-y)^2(1+x_1^- -y)}.
\end{eqnarray}

The on-shell condition for the final outgoing quark $l_q$ can be used to integrate out $d p_1^+ = p^+ d x_1$, giving $x_1=-x_0+x_B - y \; x_M +(1-y)\tilde{x}_L+x_{C1}$.
%\begin{eqnarray}
% \int \frac{d x_1} {2\pi} \delta(-x_B + x_0+{x}_1 + x_M/y - %(1-y)\tilde{x}_L- x_{C1}){e^{- i x_1 p^+ y_1^-}}=
%e^{- i (-x_0+x_B + x_{C1}+(1-y)\tilde{x}_L-x_M/y) p^+ y_1^-}.
%\end{eqnarray}
In addition, the overall momentum conservation implies, $x_1' = x_0 +x_1 - x_0' = x_B + x_{C1}+(1-y)\tilde{x}_L-y\,x_M -x_0'$, with which we can change the integral variable $x_1'$ to $x_0'$.

Now we perform the integration over the momentum fractions $x_0$ and $x_0'$ using the contour integration technique.
The $x_0$ integral can be implemented by closing the contour with a counter-clockwise semi-circle in the upper half of the $x_0$ complex plane:
\begin{eqnarray}
& &\int \frac{dx_0}{2\pi} \frac{e^{- i x_0 p^+ (y_0^- - y_1^-)}}{{(-{x}_B + x_0 - x_{M}-i\epsilon)} {(-x_B + x_0 -{x}_{M}- \tilde{x}_L -i\epsilon)}}\nonumber\\
&=& i\theta(y_1^--y_0^-) e^{-i (x_{B}+{x}_{M}+ \tilde{x}_L) p^+y_0^-}e^{i (x_{B}+{x}_{M}) p^+y_1^-}
\frac{e^{i \tilde{x}_L p^+ y_1^-}- e^{i \tilde{x}_L p^+ y_0^- }} {\tilde{x}_L}.
\end{eqnarray}
In each of the two terms above, one quark propagator is close to its mass shell (its pole is used in the contour integral), while the other propagator remains off-shell. The $\theta$-function means the quark line propagates from $y_0^-$ to $y_1^-$. For $y_0^- > y_1^-$, the integral gives zero. The integral over $x_0'$ is completely analogous, except that we close the contour
with a clockwise semi-circle in the lower half of the $x_0'$ complex plane for $y_1'^->y_0'^-=0$. A minus sign is associated with the clockwise contour integral. After the integrations over the momentum fractions $x_1$, $x_0$ and $x_0'$, the hadronic tensor takes the following form, 
\begin{eqnarray}
\label{afterphase}
W^{A\mu\nu}_{(\ref{fig2})} \!\!&=&\!\! \sum_q Q_q^{2} e^2 g^2 A C_p^A (2\pi)f_i(x_{B}+{x}_{M}+ \tilde{x}_L) \frac{1}{4 p^+ q^-}|V_{ij}|^2 {\rm Tr} [\slashed{p} \gamma^\mu (1-\gamma^5) \{\slashed{q}+ (x_B + x_M)\slashed{p}\} (1+\gamma^5) \gamma^\nu]
\nonumber\\ \!\!&\times&\!\!
\int\frac{d y}{2 \pi}\frac{1}{2 y}\int\frac{d^2\mathbf{l}_\perp}{(2\pi)^2} \int d^3\mathbf{y}_1 \int \frac{d^3\mathbf{p}_1}{(2\pi)^3} \int d^3\mathbf{y}_1' \int \frac{d^3\mathbf{p}_1'}{(2\pi)^3}e^{-i\mathbf{p}_1\cdot \mathbf{y}_1} e^{i\mathbf{p}_1'\cdot \mathbf{y}_1'}
\int dy_1^- \int d y_1'^- 
\nonumber\\\!\!&\times&\!\!
\frac{1}{N_c} {\rm Tr}\left[ T^{a_1} T^{a_1'} \right]\langle A | A_{a_1}^+(y_1) A_{a'_1}^+(y'_1) |A\rangle
\nonumber\\\!\!&\times&\!\!
\theta(y_1^--y_0^-) \theta(y_1'^--y_0'^-)e^{-i(x_{C1}+(1-y)\tilde{x}_L-(1+y)x_M) p^+(y_1^- -y_1'^-)}(e^{i \tilde{x}_L p^+ y_1^-}- e^{i \tilde{x}_L p^+ y_0^- }) (e^{-i \tilde{x}_L p^+ y_1'^-}-1)
\nonumber\\ \!\!&\times&\!\!
C_q (p^+)^3 \frac{1}{(\tilde{x}_L)^2}{\rm Tr}\left[\frac{\gamma^-}{2} (\slashed{q}_1+M) \gamma^\alpha (\slashed{\bar{q}}_1+M) \gamma^- (\slashed{l}_q+M) \gamma^- ({\slashed{\bar{q}}_1'}+M)
\gamma^\beta (\slashed{q}_1'+M)\right]G_{\alpha\beta}(l).
\end{eqnarray}
Here, we have utilized the photon on-shell condition $\delta(l^2)$ to integrate over $l^+$ and used $l^-=yq^-$ to change the variable $l^-$ into $y$. Equation~(\ref{eq:defPDF}) has also been used to convert the integral over $y_0^-$ into the light quark distribution function in a nucleon ($f_i$). Note that rigorously, the phase factor $e^{i\tilde{x}_Lp^+y_0^-}$ remaining in  Eq.~(\ref{afterphase}) should also be absorbed in the $f_i$ function. We postpone it in order to keep the general structure of $W^{A\mu\nu}_{(\ref{fig2})}$ here the same as those from the other 9 diagrams discussed in the Appendix, and also consistent with their vacuum baseline Eq.~(\ref{eq:vaccum}).

With the quark operators factorized out, we now simplify the matrix elements of the gluon vector potentials in the nucleus state. Considering that the whole nucleus is in a color singlet state, the expectation of the gluon field operators can be expressed as follows:
\begin{eqnarray}
\langle A | A_{a_1}^+(y_1) A_{a_1'}^+(y_1') |A\rangle
= \frac{1}{d(R)} {\rm Tr} [T_{a_1}(R) T_{a_1'}(R)] \langle A | A^+(y_1) A^+(y_1') |A\rangle
= \frac{\delta_{a_1 a_1'} C_2(R)}{N_c^2-1} \langle A| A^+(y_1) A^+(y_1') | A\rangle,
\end{eqnarray}
where $d(R)$ and $C_2(R)$ are the dimension and the quadratic Casimir of the representation $R$ of the $SU(3)$ color group. For a gluon exchanged from a quark line, $R$ is the fundamental representation with $d(R) = 3$ and $C_2(R) = C_F = 4/3$. For a gluon exchanged from a gluon line, $R$ is the adjoint representation with $d(R) = 8$ and $C_2(R) = C_A = 3$. Since the two exchanged gluons, one in the amplitude and one in the complex conjugate, carry the same color ($\delta_{a_1 a_1'}$), one may evaluate the color factor of the hadronic tensor $W^{A\mu\nu}_{(\ref{fig2})}$ as
\begin{eqnarray}
\frac{\delta_{a_1 a_1'}}{N_c} {\rm Tr}[T^{a_1}T^{a_1'}] = C_F .
\end{eqnarray}

In the limit of very high energy, the nucleus may be approximated by a weakly interacting homogenous gas of nucleons. This allows us to change the variables $(y_1, y_1')\rightarrow (Y_1, \delta y_1)$ with
\begin{eqnarray}
Y_1 = \frac{1}{2}(y_1 + y_1'), \;\;\;\; \delta y_1 = y_1 - y_1',
\end{eqnarray}
and assume translational invariance of the correlation function with respect to $Y_1$ as
$\langle A| A^+(y_1) A^+(y'_1) | A\rangle \approx \langle A| A^+(\delta{y_1}) A^+(0) | A\rangle$.
The integrals of the related phase factors yield:
\begin{eqnarray}
\int d^3\mathbf{y}_1 \int d^3\mathbf{y}_1' e^{-i\mathbf{p}_1 \cdot \mathbf{y}_1} e^{i\mathbf{p}_1'\cdot \mathbf{y}_1'}
= (2\pi)^3 \delta(\mathbf{p}_1 - \mathbf{p}_1') \int d^3\delta \mathbf{y}_1 e^{-i(\mathbf{p}_1+\mathbf{p}_1')\cdot \frac{\delta\mathbf{y}_1}{2}}.
\end{eqnarray}
In the expression above, the three-$\delta$-function represents the fact that the two exchanged gluons carry the same momentum ($\mathbf{p}_1' = \mathbf{p}_1$), which further implies $\mathbf{p}_0' = \mathbf{p}_0=0$.
This $\delta$-functions is also utilized to complete the integral over the momentum $\mathbf{p}'_1$.

We denote the phase factors of Eq.~(\ref{afterphase})---the second-to-last line---as $S_{(\ref{fig2})}$ and re-write it as follows,
\begin{eqnarray}
\label{eq:phase}
S_{(\ref{fig2})} &=& e^{-i(x_{C1}+(1-y)\tilde{x}_L-(1+y)x_M) p^+ \delta y_1^-}(e^{i \tilde{x}_L p^+ (Y_1^- + \delta y_1^-/2)}- e^{i \tilde{x}_L p^+ y_0^- }) (e^{-i \tilde{x}_L p^+ (Y_1^- - \delta y_1^-/2)}-1)\nonumber\\
&=& e^{-i(x_{C1}+(1-y)\tilde{x}_L-(1+y)x_M) p^+ \delta y_1^-}e^{i \tilde{x}_L p^+ y_0^- }(e^{i \tilde{x}_L p^+ (Y_1^- + \delta y_1^-/2- y_0^-)}- 1) (e^{-i \tilde{x}_L p^+ (Y_1^- - \delta y_1^-/2)}-1).%\\
%&=& e^{-i(x_{C1}+(1-y)\tilde{x}_L-(1+y)x_M) p^+ \delta y_1^-}e^{i \tilde{x}_L p^+ y_0^- }{\color{blue}(e^{i \tilde{x}_L p^+ Y_1^- }- 1) (e^{-i \tilde{x}_L p^+ Y_1^- }-1)}
\end{eqnarray}
Note that $Y_1^-$ represents the location of the rescattering gluon, which spans over the size of the nucleus, while $y_0^-$ and $\delta y_1^-$ are confined within the size of one nucleon due to the short range correlations of the quark and gluon fields.
Thus, $y_0^-$ and $\delta y_1^-$ are much smaller than $Y_1^-$ and can be neglected in the last two terms of the expression above. As discussed earlier, the second term ($e^{i \tilde{x}_L p^+ y_0^- }$) should be absorbed in the $f_i$ in Eq.~(\ref{afterphase}). However, since $\tilde{x}_L\ll x_B$ is assumed, this term can be neglected. In addition, we assume $\delta y_1^-$ is sufficiently small that the first term can be neglected as well. In the end, Eq.~(\ref{eq:phase}) is approximated as
\begin{eqnarray}
S_{(\ref{fig2})} \approx 2-2 \cos(\tilde{x}_L p^+ Y_1^-).
\end{eqnarray}

The trace part of Eq.~(\ref{afterphase}) is evaluated as
\begin{eqnarray}
& &\!\! {\rm Tr}\left[\frac{\gamma^-}{2} (\slashed{q}_1+M) \gamma^\alpha (\slashed{\bar{q}}_1+M) \gamma^- (\slashed{l}_q+M) \gamma^- ({\slashed{\bar{q}}_1'}+M)
\gamma^\beta (\slashed{q}_1'+M)\right]G_{\alpha\beta}(l)\nonumber\\
\!\!&=&\!\! 16 (q^-)^3 (1+x_1^- -y)\frac{P(y)}{y}\left(l_\perp^2+\frac{y^4}{1+(1-y)^2}M^2\right).
\end{eqnarray}
Therefore, one may obtain the hard part of the matrix element---the last line of Eq.~(\ref{afterphase}), denoted as $T_{(\ref{fig2})}$---as
\begin{eqnarray}
T_{(\ref{fig2})} \!\!&=&\!\! {2yP(y)}\frac{l_\perp^2+\frac{y^4}{1+(1-y)^2}M^2}{(l_\perp^2+y^2 M^2)^2}.
\end{eqnarray}

With all the simplifications above, the hadronic tensor reads:
\begin{eqnarray}
W^{A\mu\nu}_{(\ref{fig2})} \!\!&=&\!\! \sum_q Q_q^2  A C_p^A (2\pi)f_i(x_{B}+{x}_{M}+ \tilde{x}_L) \frac{1}{4 p^+ q^-}|V_{ij}|^2 {\rm Tr} [\slashed{p} \gamma^\mu (1-\gamma^5) \{\slashed{q}+ (x_B + x_M)\slashed{p}\} (1+\gamma^5) \gamma^\nu]
\nonumber\\ \!\!&\times&\!\!
 \int d Y_1^-\int \frac{d^3\mathbf{p}_1}{(2\pi)^3}\int d \delta y_1^- \int d^3 \delta \mathbf{y}_{1} e^{-i \mathbf{p}_{1} \cdot \delta \mathbf{y}_{1}}\left( g^2 \frac{C_F C_2(R)}{N_c^2-1}\right)\langle A | A^+(\delta y_1^-, \delta \mathbf{y}_{1}) A^+(0) | A\rangle
 \nonumber\\ \!\!&\times&\!\!
 \frac{\alpha_e}{2 \pi}\int dy P(y) \int {dl_\perp^2}
\left[2-2 \cos\left( \frac{Y_1^-}{\tilde{\tau}_{\rm{form}}^-}\right)\right]\frac{l_\perp^2+\frac{y^4}{1+(1-y)^2}M^2}{(l_\perp^2+y^2 M^2)^2}.
\end{eqnarray}
The $\theta$-functions in Eq.~(\ref{afterphase}) are included in the range of the integral over $Y_1^-$, $(0,L)$, with $L$ the path length of the hard quark inside the nuclear medium.
One can define the distribution function of the exchanged momentum between the hard quark and the medium in light-cone gauge as 
\begin{eqnarray}
\label{eq:defD}
\mathcal{D}(\,p_1^-,\, \mathbf{p}_{1 \perp})=\int d \delta y_1^- \int d^3 \delta \mathbf{y}_{1} e^{-i \mathbf{p}_{1} \cdot \delta \mathbf{y}_{1}}\left( g^2 \frac{C_F C_2(R)}{N_c^2-1}\right)\langle A | A^+(\delta y_1^-, \delta \mathbf{y}_{1}) A^+(0) |A\rangle,\;\;\;\;\;\;\;\;\;
\end{eqnarray}
with which the medium-induced single photon bremsstrahlung spectrum from a hard quark for Fig.~\ref{fig2} can be written as
\begin{eqnarray}
\frac{d N^\mathrm{med}_{\gamma,\,(\ref{fig2})}}{d y d {l}_\perp^2} &=&
\frac{\alpha_e}{2 \pi}{P(y)}\int d Y_1^- \int \frac{d^3\mathbf{p}_1}{(2\pi)^3}\mathcal{D}(\,p_1^-,\, \mathbf{p}_{1 \perp}) \left[2-2 \cos\left( \frac{Y_1^-}{\tilde{\tau}_{\rm{form}}^-}\right)\right]\frac{l_\perp^2+\frac{y^4}{1+(1-y)^2}M^2}{(l_\perp^2+y^2 M^2)^2}.
\end{eqnarray}

There are $9$ other diagrams contributing to the photon bremsstrahlung process at the next-to-leading twist (twist-four) level, or the single photon emission induced by a single scattering between the hard quark and the nuclear medium. Their calculations are analogous. We list their main results in the Appendix. After summing over the results of all the $10$ diagrams, the medium-induced photon bremsstrahlung spectrum is written as
\begin{eqnarray}
\label{heavyquark1}
\frac{d N^{med}_{\gamma}}{d y d {l}_\perp^2} &=&
\frac{\alpha_e}{2 \pi}{P(y)}\int d Y_1^- \int \frac{d^3\mathbf{p}_1}{(2\pi)^3}\mathcal{D}(\,p_1^-,\, \mathbf{p}_{1\, \perp})\Biggl\{\left[2-2 \cos\left( \frac{Y_1^-}{\tilde{\tau}_{\rm{form}}^-}\right)\right]
\nonumber\\ 
& & \times \left[\frac{l_\perp^2+\frac{y^4}{1+(1-y)^2}M^2}{(l_\perp^2+y^2 M^2)^2}
-\frac{1+\left(1 - y \right)\left(1-\frac{y}{1+ x_1^-}\right)}{1+(1-y)^2} \frac{\mathbf{l}_{\perp} \cdot \left(\mathbf{l}_{\perp} - \frac{y}{1+ x_1^-} \mathbf{p}_{1\perp}\right)+\frac{y^2\left(\frac{y}{1+x_1^-}\right)^2 M^2}{1+(1-y)\left(1-\frac{y}{1+x_1^-}\right)}}{(l_\perp^2 + y^2 M^2)
\left[\left(\mathbf{l}_{\perp} - \frac{y}{1+ x_1^-} \mathbf{p}_{1\perp}\right)^2+\left(\frac{y}{1+ x_1^-}\right)^2 M^2\right]}\right]\nonumber\\
& &\left. + \frac{1+\left(1-\frac{y}{1+x_1^-}\right)^2}{1+(1-y)^2}
\frac{\left(\mathbf{l}_\perp-\frac{y}{1+x_1^-}\mathbf{p}_{1 \perp}\right)^2+\frac{\left(\frac{y}{1+ x_1^-}\right)^4 M^2}{1+\left(1-\frac{y}{1+ x_1^-}\right)^2}}{
\left[\left(\mathbf{l}_\perp-\frac{y}{1+x_1^-}\mathbf{p}_{1 \perp}\right)^2+ \left(\frac{y}{1+ x_1^-}\right)^2 M^2\right]^2}-\frac{l_\perp^2+\frac{y^4}{1+(1-y)^2}M^2}{(l_\perp^2+y^2 M^2)^2}\right\}.
\end{eqnarray}
This expression is general in the sense that the information about the dense nuclear medium traversed by the hard quark is encoded in the distribution function of the exchanged momentum $\mathcal{D}(\,p_1^-,\, \mathbf{p}_{1 \perp})$. As discussed in Appendix B of Ref.~\cite{Zhang:2018nie}, this distribution function can be related to the differential rate of elastic scatterings between the hard parton and the nuclear medium:
\begin{eqnarray}
\mathcal{D}(\,p_1^-,\, \mathbf{p}_{1 \perp}) %= (2\pi)^3 \frac{d\Gamma_{\rm el}}{d^3p_1}
= (2\pi)^3 \frac{dP_{\rm el}}{d p_1^- d^2\mathbf{p}_{1\perp} d Y_1^-}.
\end{eqnarray}
As we work beyond the collinear expansion approximation where $\langle p_{1\perp}^2\rangle \ll l_\perp^2$ is assumed, the medium-induced photon emission spectrum is directly governed by the full distribution of the differential elastic scattering rate---${dP_{\rm el}}/{(d p_1^- d^2\mathbf{p}_{1\perp} d Y_1^-)}$. Contributions from both transverse and longitudinal momentum exchanges between the hard quark and the medium are included. In turn, one may use the photon emission spectrum, or the jet energy loss, to probe the properties of the nuclear medium encoded in the $\mathcal{D}(\,p_1^-,\, \mathbf{p}_{1 \perp})$ function. 

With the quark mass ($M$) set to zero, Eq.~(\ref{heavyquark1}) can be reduced to
\begin{eqnarray}
\label{lightquark1}
\frac{d N^\mathrm{med}_{\gamma}}{d y d {l}_\perp^2} &=&
\frac{\alpha_e}{2 \pi}\frac{P(y)}{{l}_\perp^2}\int d Y_1^- \int \frac{d^3\mathbf{p}_1}{(2\pi)^3}\mathcal{D}(\,p_1^-,\, \mathbf{p}_{1 \perp})
\nonumber\\&\times& \left\{\left[2-2 \cos\left( \frac{Y_1^-}{{\tau}_{\rm{form}}^-}\right)\right]\left[1-\frac{1+(1-y)\left(1-\frac{y}{1+x_1^-}\right)}{1+(1-y)^2}
\frac{\mathbf{l}_\perp \cdot \left(\mathbf{l}_\perp -\frac{y}{1+x_1^-}\mathbf{p}_{1\perp} \right) }{\left(\mathbf{l}_\perp -\frac{y}{1+x_1^-}\mathbf{p}_{1\perp} \right)^2}\right]
\right.\nonumber\\&+&
\left.\frac{1+\left(1-\frac{y}{1+x_1^-}\right)^2}{1+(1-y)^2}\frac{{l}_\perp^2 }{\left(\mathbf{l}_\perp -\frac{y}{1+x_1^-}\mathbf{p}_{1\perp} \right)^2}-1\right\},
\end{eqnarray}
in which the formation time of photon emission from a massless quark reads: $\tilde{\tau}_{\rm{form}}^-=2q^-y(1-y)/l_\perp^2$. Compared to Ref.~\cite{Zhang:2016avg}, this provides the medium-induced photon spectrum from a hard light quark beyond the collinear expansion approximation.

As an example, we assume the nuclear medium traversed by the hard quark is composed of heavy static scattering centers and adopt a static Debye screened potential for the exchanged gluon field in the light-cone gauge as~\cite{Sievert:2018imd},
\begin{eqnarray}
\label{yukawa potential2}
A^{\mu}(\mathbf{p})=g^{\mu - } (2 \pi)\delta(p^-)\frac{- g}{\mathbf{p}_{\perp}^2+\mu_D^2},
\end{eqnarray}
where $\mu_D$ is the mass of an exchanged gluon.
Recall that the color factor of the gluon field has already been factored out. In the static potential above, the $A^+$ component is nonzero and the $p^-$ component is set to zero, since we work in the high-energy limit of the incoming nucleus and use the light-cone gauge throughout our calculation.
% This indicates that we have used the light-cone coordinate throughout our calculation. In the high energy limit, we only keep the dominant $(+)$ component of the scattered gluon field. 
For a hard parton scattering via this static potential, only transverse momentum is exchanged between the hard parton and the nuclear medium. To include the contribution from longitudinal momentum (and/or energy) exchange between the hard parton and the medium, one may consider dynamical constituents for the dense nuclear medium~\cite{Arnold:2001ba,Arnold:2002ja,Djordjevic:2007at,Djordjevic:2008iz,Karmakar:2024fkn}. By performing the Fourier transformation, we may write the correlation function of the exchanged gluon fields as
\begin{eqnarray}
\label{AA IN CARTESIAN2}
\langle A | A^\mu(\delta \mathbf{y}_1) A^\nu(0)| A \rangle &=&
\delta_+^\mu \delta_+^\nu \rho^-(\delta \mathbf{y}_{1}) \delta(\delta y_1^-)\int \frac{d^2 \mathbf{p}_\perp }{(2 \pi)^2}e^{i \mathbf{p}_\perp \cdot {\delta \mathbf{y}_{1 \perp}}} \frac{g^2}{(\mathbf{p}_\perp^2 + \mu_D^2)^2},\;\;\;\;\;\;
\end{eqnarray}
where $\rho^-$ is the light-cone density of the medium constituents (scattering centers) that the hard jet parton interacts with.

Using the definition of the distribution function of the exchanged gluon, Eq.~(\ref{eq:defD}), and assuming the medium density is a constant $\rho^-$, one can obtain
\begin{equation}
\label{eq:defDperp}
\mathcal{D}(p_1^-,\,\mathbf{p}_{1 \perp})= (2\pi)\delta(p_1^-) (2 \pi)^2 \rho^- \frac{ C_F C_2(R)}{N_c^2-1} \frac{4\alpha_s^2}{(\mathbf{p}_{1\perp}^2 + \mu_D^2)^2} \equiv (2\pi)\delta(p_1^-) \mathcal{D}_{\perp}(\mathbf{p}_{1 \perp}).
\end{equation}
Note that the differential elastic cross section for light quark scattering with the static Debye screened potential in light-cone gauge is
\begin{eqnarray}
\frac{d\sigma_{\rm el}}{d^2 \mathbf{p}_{1\perp}} = \frac{ C_F C_2(R) }{N_c^2-1}\frac{ |g A^+(\mathbf{p}_{1\perp}) |^2}{4\pi^2} = \frac{ C_F C_2(R)}{N_c^2-1} \frac{4\alpha_s^2}{(\mathbf{p}_{1\perp}^2 + \mu_D^2)^2}.\;\;\;\;\;\;\;
\end{eqnarray}
Therefore, we have
%Using the above expressions, the distribution function of the rescattered momentum in light-cone gauge may be written as,
%\begin{eqnarray}
%\mathcal{D}(p_1^-,\,\mathbf{p}_{1 \perp})&=& (2\pi)\delta(p_1^-) (2 \pi)^2 \rho^-\frac{d\sigma_{\rm el}}{d^2 \mathbf{p}_{1\,\perp}} = (2\pi)\delta(p_1^-) \mathcal{D}_{\perp}(\mathbf{p}_{1 \perp}),
%\end{eqnarray}
%here for convenience, the distribution function $\mathcal{D}_{\perp}(\mathbf{p}_{1 \perp})$ for transverse momentum exchange have also been defined as,
\begin{equation}
\mathcal{D}_{\perp}(\mathbf{p}_{1 \perp})  =  (2 \pi)^2 \rho^-\frac{d\sigma_{\rm el}}{d^2 \mathbf{p}_{1\perp}}
= (2 \pi)^2 \frac{d P_{\rm el}}{d^2 \mathbf{p}_{1\perp}dY_1^-},
\end{equation}
where ${dP_{\rm el}}/{(d^2 \mathbf{p}_{1\perp}d Y_1^-)}$ is the differential elastic scattering rate of the hard parton inside the nuclear medium.
The distribution function of the exchanged gluon can be related to the (light-cone) jet transport coefficient $\hat{q}$ via
\begin{eqnarray}
\hat{q} \equiv \frac{d\langle p_{1\perp}^2\rangle}{dL^-} = \int \frac{d p_1^- d^2\mathbf{p}_{1\perp}}{(2\pi)^3} \mathbf{p}_{1 \perp}^2 \mathcal{D}(p_1^-,\,\mathbf{p}_{1 \perp}) =\int \frac{d^2\mathbf{p}_{1\perp}}{(2\pi)^2} \mathbf{p}_{1 \perp}^2 \mathcal{D}_\perp(\mathbf{p}_{1 \perp}) =\int {d^2\mathbf{p}_{1\perp}} \mathbf{p}_{1 \perp}^2 \rho^-\frac{d\sigma_{\rm el}}{d^2 \mathbf{p}_{1\perp}}.
\end{eqnarray}

By substituting Eq.~(\ref{eq:defDperp}) into Eq.~(\ref{heavyquark1}), we have 
\begin{eqnarray}
\label{heavyquark2}
\frac{d N^\mathrm{med}_{\gamma}}{d y d{l}_\perp^2} &=&
\frac{\alpha_e}{2 \pi}{P(y)}\int d Y_1^- \int \frac{d^2\mathbf{p}_{1\perp}}{(2\pi)^2}\mathcal{D}_{\perp}(\mathbf{p}_{1 \perp})
\nonumber\\ 
&\times& \left\{\left[2-2 \cos\left( \frac{Y_1^-}{{\tilde{\tau}}_{\rm{form}}^-}\right)\right]\left[\frac{l_\perp^2+\frac{y^4}{1+(1-y)^2}M^2}{(l_\perp^2+y^2 M^2)^2}-\frac{\mathbf{l}_{\perp} \cdot \left(\mathbf{l}_{\perp} - {y} \mathbf{p}_{1\perp}\right)+\frac{y^4 M^2}{1+(1-y)^2}}{(l_\perp^2 + y^2 M^2)
\left[\left(\mathbf{l}_{\perp} - y \mathbf{p}_{1\perp}\right)^2+y^2 M^2\right]}\right]
\right.\nonumber\\
&+& \left. \frac{\left(\mathbf{l}_\perp-y\mathbf{p}_{1 \perp}\right)^2+\frac{y^4 M^2}{1+\left(1-y\right)^2}}{
\left[\left(\mathbf{l}_\perp-y\mathbf{p}_{1 \perp}\right)^2+ y^2 M^2\right]^2}-\frac{l_\perp^2+\frac{y^4}{1+(1-y)^2}M^2}{(l_\perp^2+y^2 M^2)^2}\right\}.
\end{eqnarray}
This is the spectrum of single scattering induced single photon emitted from a massive hard quark when the static Debye screened potential model is used to evaluate the correlation function of the exchanged gluon fields in the light-cone gauge. Working beyond the collinear expansion approximation, we find that the single photon bremsstrahlung spectrum induced by single rescattering with only transverse momentum exchange is determined by the distribution of the transverse momentum exchange---$\mathcal{D}_\perp(\mathbf{p}_{1\perp})$, which is proportional to the differential elastic scattering rate---$dP_\mathrm{el}/(d^2 p_{1\perp} dY_1^-)$---experienced by the hard quark inside the dense nuclear medium.

By setting the quark mass to zero, Eq.~(\ref{heavyquark2}) is reduced to the medium-induced photon spectrum emitted from a hard light quark within the static Debye screened potential model:
%including only transverse momentum exchanges with the medium constituents when we neglect the mass $M$ of the quark jet, 
\begin{eqnarray}
\label{lightquark2}
\frac{d N^\mathrm{med}_{\gamma}}{d y d{l}_\perp^2} &=&
\frac{\alpha_e}{2 \pi}\frac{P(y)}{l_\perp^2}\int d Y_1^- \int \frac{d^2\mathbf{p}_{1\perp}}{(2\pi)^2}\mathcal{D}_{\perp}(\mathbf{p}_{1 \perp})\nonumber\\
&\times& \left\{\left[2-2 \cos\left( \frac{Y_1^-}{{{\tau}}_{\rm{form}}^-}\right)\right]
\left[1-
\frac{\mathbf{l}_\perp \cdot \left(\mathbf{l}_\perp -y \mathbf{p}_{1\perp} \right) }{\left(\mathbf{l}_\perp -y \mathbf{p}_{1\perp} \right)^2}\right]+
\frac{{l}_\perp^2}{\left(\mathbf{l}_\perp - y \mathbf{p}_{1\perp} \right)^2}-1\right\}.
\end{eqnarray}

%\end{widetext}

\section{Summary}
\label{sec:IV}

Within the framework of deep inelastic scattering, we have studied the process of single scattering induced single photon emission from a hard heavy or light quark inside a dense nuclear medium. The charged current interaction channel is considered, in which a $W$ boson emitted by a projectile electron strikes a quark inside a target nucleus and converts it into a hard (either heavy or light flavor) quark. We take into account one rescattering between the hard quark and the nuclear medium and evaluate the spectrum of the bremsstrahlung photon at the next-to-leading twist (twist-four) level. Calculations of all (10) diagrams, corresponding to different orders between quark rescattering and photon emission in the amplitude and its complex conjugate, are included. The full result for a heavy quark is given by Eq.~(\ref{heavyquark1}) and its simplification to a light quark is given by Eq.~(\ref{lightquark1}). Beyond the collinear expansion approximation for the momentum exchange of the rescattering, we show that the photon spectrum is governed by a full distribution of the exchanged gluon momentum, characterized by $\mathcal{D}(p_1^-,\mathbf{p}_{1\perp})$ in Eq.~(\ref{eq:defD}), which is equivalent to the differential elastic scattering rate ${dP_{\rm el}}/{(d p_1^- d^2\mathbf{p}_{1\perp} d Y_1^-)}$. In general, both longitudinal ($p_1^-$) and transverse momentum ($\mathbf{p}_{1\perp}$) exchanges are included. As a special case, we model the dense nuclear medium with heavy static scattering centers and adopt a static Debye screened potential for the exchanged gluon field. This suppresses the longitudinal momentum exchange between the hard quark and the medium, making the medium-induced photon spectrum solely depend on the transverse momentum distribution of the exchanged gluon, characterized by $\mathcal{D}(\mathbf{p}_{1\perp})$ or ${dP_{\rm el}}/{(d^2\mathbf{p}_{1\perp} d Y_1^-)}$. Using this model, our result for heavy quark is reduced to Eq.~(\ref{heavyquark2}), and its massless version for light quark is given by Eq.~(\ref{lightquark2}).

Together with the gluon bremsstrahlung process from a massive quark traversing a dense nuclear medium beyond the collinear expansion approximation~\cite{Zhang:2018kkn,Zhang:2018nie}, this series of work will provide more reliable theoretical inputs to phenomenological studies on photon production and jet quenching in relativistic heavy-ion collisions. We will incorporate these updated bremsstrahlung photon and gluon spectra in our realistic simulation of jet interactions with the QGP in our upcoming efforts, improving our understanding of the mass effect on both jet-induced photon production and jet energy loss inside the QGP.

\section*{Acknowledgments}

This work is supported by the National Natural Science Foundation of China (NSFC) under Grant Nos.~12005056, 12575146, 12175122, 2021-867, 12321005, 12225503, 12435009, and 12275104. This work is also supported in part by the National
Key Research and Development Program of China under Contract
No. 2022YFA1604900. 

%\begin{widetext}
%
\section*{Appendix}
\begin{figure}[thb]
\centering
\includegraphics[width=0.45\linewidth]{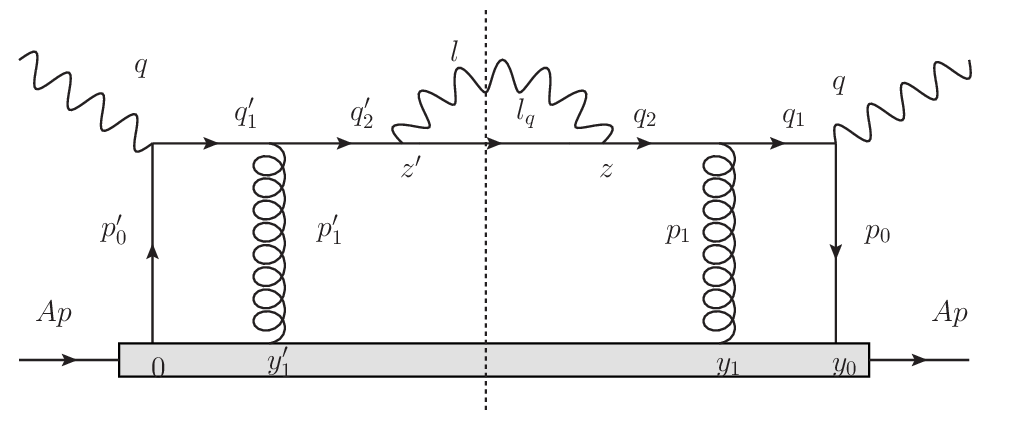}
\caption{A central-cut diagram: one rescattering on the quark
before photon bremsstrahlung in both the amplitude and the complex
conjugate.} \label{fig3}
\end{figure}

\begin{figure}[thb]
\centering
\includegraphics[width=0.95\linewidth]{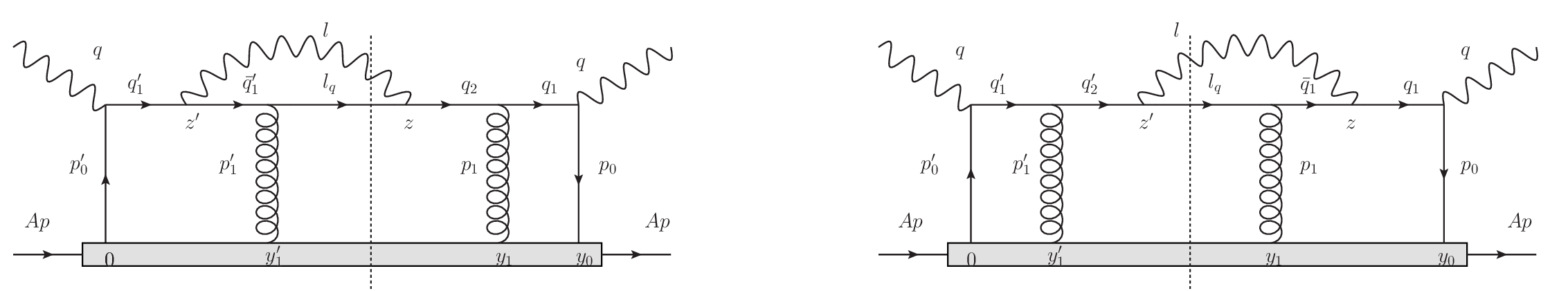}
\caption{Two central-cut diagrams: one rescattering on the quark before photon bremsstrahlung in the amplitude and one rescattering on the quark
after photon bremsstrahlung in the complex conjugate (a) or vice versa (b).} \label{fig4}
\end{figure}

\begin{figure}[thb]
\centering
\includegraphics[width=0.95\linewidth]{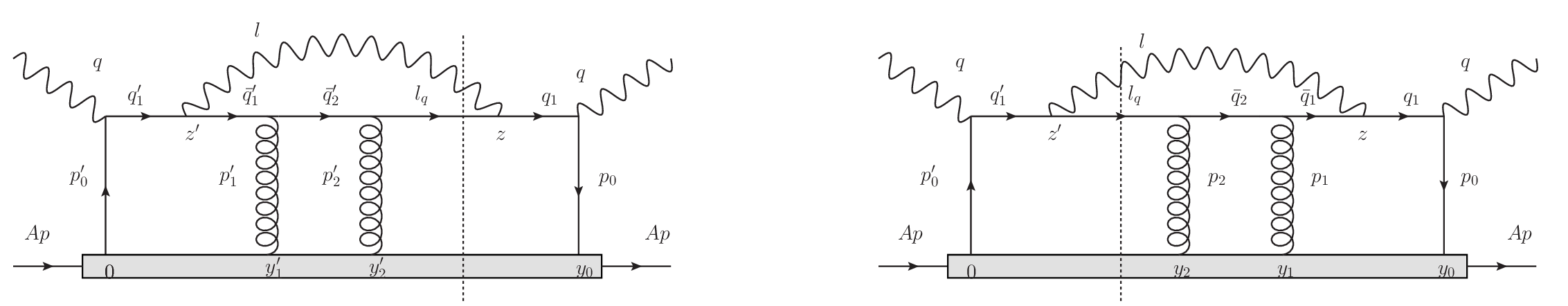}
\caption{ Two non-central-cut diagrams: two rescatterings on the quark after photon bremsstrahlung in the amplitude and zero rescattering in the complex conjugate (a) or vice versa (b).} \label{fig5}
\end{figure}

\begin{figure}[thb]
\centering
\includegraphics[width=0.95\linewidth]{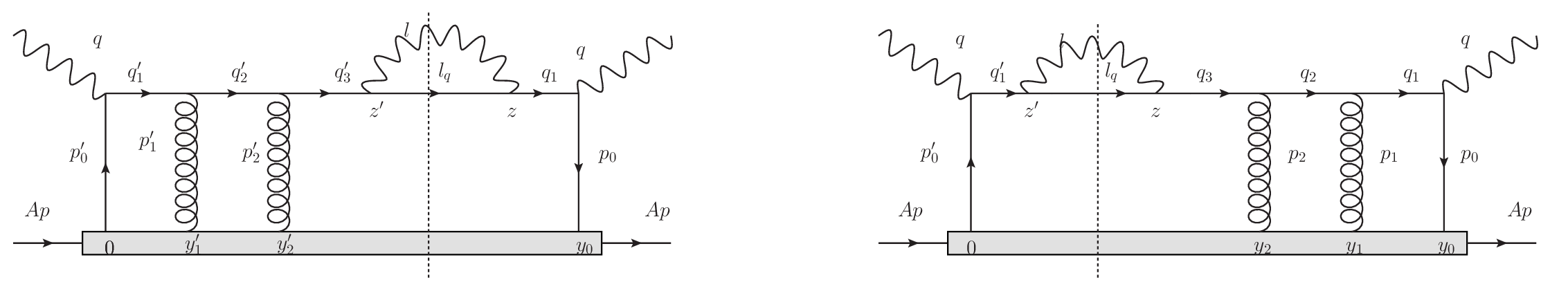}
\caption{ Two non-central-cut diagrams: two rescatterings on the quark before photon bremsstrahlung in the amplitude and zero rescattering in the complex conjugate (a) or vice versa (b).} \label{fig6}
\end{figure}

\begin{figure}[thb]
\centering
\includegraphics[width=0.96\linewidth]{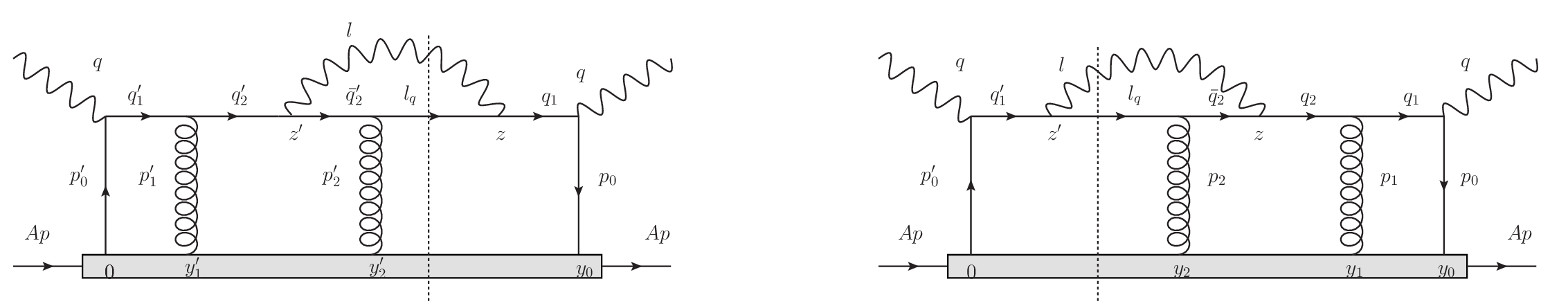}
\caption{Two non-central-cut diagrams: two rescatterings (with one on the quark before photon bremsstrahlung and one on the quark after photon bremsstrahlung) in the amplitude and zero rescattering in the complex conjugate (a) or vice versa (b).} \label{fig7}
\end{figure}

In this Appendix, we present the main results of the hadronic tensor obtained from the other 9 diagrams, as shown in Figs.~\ref{fig3}-\ref{fig7}. Among them, Figs.~\ref{fig3}-\ref{fig4} include 3 central-cut diagrams, and Figs.~\ref{fig5}-\ref{fig7} include 6 non-central-cut diagrams.

The phase factor for Fig.~\ref{fig3} reads:
\begin{eqnarray}
S_{(\ref{fig3})}&=&e^{i \tilde{x}_L p^+ y_0^-}e^{-i[(1-y)\tilde{x}_L+x_{C1}-(1+y)x_M]p^+(y_1^--y_1'^-)}
\nonumber\\&\approx&e^{i \tilde{x}_L p^+ y_0^-}e^{-i[(1-y)\tilde{x}_L+x_{C1}-(1+y)x_M]p^+\delta y_1^-}
\approx 1.
\end{eqnarray}
The hard part of the matrix element for Fig.~\ref{fig3} reads:
\begin{eqnarray}
{T}_{(\ref{fig3})} &=& {2yP(y)} \frac{1+\left(1-\frac{y}{1+x_1^-}\right)^2}{1+(1-y)^2}
\frac{\left(\mathbf{l}_\perp-\frac{y}{1+x_1^-}\mathbf{p}_{1 \perp}\right)^2+\frac{\left(\frac{y}{1+ x_1^-}\right)^4 M^2}{1+\left(1-\frac{y}{1+ x_1^-}\right)^2}}{
\left[\left(\mathbf{l}_\perp-\frac{y}{1+x_1^-}\mathbf{p}_{1 \perp}\right)^2+ \left(\frac{y}{1+ x_1^-}\right)^2 M^2\right]^2}.
\end{eqnarray}

The phase factor for Fig.~\ref{fig4} reads:
\begin{eqnarray}
S_{(\ref{fig4},a)}&=& e^{i \tilde{x}_L p^+ y_0^-}e^{-i[(1-y)\tilde{x}_L+x_{C1}-(1+y)x_M]p^+(y_1^- -y_1'^-)} (e^{-i \tilde{x}_L p^+ y_1'^-}-1)
\nonumber\\
&\approx& e^{i \tilde{x}_L p^+ y_0^-}e^{-i[(1-y)\tilde{x}_L+x_{C1}-(1+y)x_M]p^+\delta y_1^- } (e^{-i \tilde{x}_L p^+ (Y_1^--\delta y_1^-/2)}-1)
\approx e^{-i \tilde{x}_L p^+ Y_1^-}-1,
\nonumber\\
S_{(\ref{fig4},b)}&=& e^{i \tilde{x}_L p^+ y_0^-}e^{-i[(1-y)\tilde{x}_L+x_{C1}-(1+y)x_M]p^+(y_1^- -y_1'^-)}\left(e^{i\tilde{x}_L p^+ y_1^-} - e^{-i \tilde{x}_L p^+ y_0^-}\right) \nonumber\\
&\approx& e^{-i[(1-y)\tilde{x}_L+x_{C1}-(1+y)x_M]p^+\delta y_1^- } (e^{i \tilde{x}_L p^+ (Y_1^-+\delta y_1^-/2)}-e^{-i \tilde{x}_L p^+ y_0^-})
\approx e^{i \tilde{x}_L p^+ Y_1^-}-1,
\nonumber\\
S_{(\ref{fig4})}&=&S_{(\ref{fig4},a)}+S_{(\ref{fig4},b)}\approx 2  \cos(\tilde{x}_L p^+ Y_1^-) -2.
\end{eqnarray}
The hard part of the matrix element for Fig.~\ref{fig4} reads:
\begin{eqnarray}
{T}_{(\ref{fig4},a)} ={T}_{(\ref{fig4},b)} &=& {2yP(y)}
\frac{1+\left(1 - y \right)\left(1-\frac{y}{1+ x_1^-}\right)}{1+(1-y)^2} \frac{\mathbf{l}_{\perp} \cdot \left(\mathbf{l}_{\perp} - \frac{y}{1+ x_1^-} \mathbf{p}_{1\perp}\right)+\frac{y^2\left(\frac{y}{1+x_1^-}\right)^2 M^2}{1+(1-y)\left(1-\frac{y}{1+x_1^-}\right)}}{(l_\perp^2 + y^2 M^2)
\left[\left(\mathbf{l}_{\perp} - \frac{y}{1+ x_1^-} \mathbf{p}_{1\perp}\right)^2+\left(\frac{y}{1+ x_1^-}\right)^2 M^2\right]}.
\end{eqnarray}

The phase factor for Fig.~\ref{fig5} reads:
\begin{eqnarray}
S_{(\ref{fig5},a)}&=& - \frac{1}{2} e^{i[x_{C1}-y\tilde{x}_L-(1+y)x_M]p^+(y_2'^- -y_1'^-)}(1 - e^{i \tilde{x}_L p^+ y_1'^-})
\nonumber\\ &\approx& - \frac{1}{2} e^{i[x_{C1}-y\tilde{x}_L-(1+y)x_M]p^+\delta y_1^-}(1- e^{i \tilde{x}_L p^+ (Y_1^--\delta y_1^-/2)})
\approx - \frac{1}{2}(1-e^{i \tilde{x}_L p^+ Y_1^-}),
\nonumber\\
S_{(\ref{fig5},b)}&=& - \frac{1}{2} e^{-i \tilde{x}_L p^+ y_0^-} e^{-i[x_{C1}-y\tilde{x}_L-(1+y)x_M]p^+(y_2^- -y_1^-)}(e^{-i \tilde{x}_L p^+ y_0^-} - e^{-i \tilde{x}_L p^+ y_1^-})
\nonumber\\&\approx&- \frac{1}{2} e^{-i \tilde{x}_L p^+ y_0^-}e^{-i[x_{C1}-y\tilde{x}_L-(1+y)x_M]p^+\delta y_1^-}(e^{-i \tilde{x}_L p^+ y_0^-}- e^{-i \tilde{x}_L p^+ (Y_1^--\delta y_1^-/2)})
\approx - \frac{1}{2}(1-e^{-i \tilde{x}_L p^+ Y_1^-}),
\nonumber\\
S_{(\ref{fig5})}&=&S_{(\ref{fig5},a)}+S_{(\ref{fig5},b)}\approx \cos(\tilde{x}_L p^+ Y_1^-) -1.
\end{eqnarray}
The hard part of the matrix element for Fig.~\ref{fig5} reads:
\begin{eqnarray}
{T}_{(\ref{fig5},a)} ={T}_{(\ref{fig5},b)} &=& {2yP(y)}\frac{l_\perp^2+\frac{y^4}{1+(1-y)^2}M^2}{(l_\perp^2+y^2 M^2)^2}.
\end{eqnarray}

The phase factor for Fig.~\ref{fig6} reads:
\begin{eqnarray}
S_{(\ref{fig6},a)}& =&- \frac{1}{2}e^{i (x_{D1}'-x_M) p^+ (y_1'^--y_2'^-)}e^{i \tilde{x}_{L} p^+ y_2'^-}
\nonumber\\&\approx& - \frac{1}{2} e^{-i (x_{D1}-x_M) p^+ \delta y_1^-} e^{i \tilde{x}_L p^+ (Y_1^-+\delta y_1^-/2)}\approx - \frac{1}{2} e^{i \tilde{x}_L p^+ Y_1^-},
\nonumber\\
S_{(\ref{fig6},b)}&=&- \frac{1}{2}e^{i \tilde{x}_{L} p^+ y_0^-}e^{-i (x_{D1}-x_M) p^+ (y_1^--y_2^-)}e^{- i \tilde{x}_{L} p^+ y_2^-}
\nonumber\\&\approx& - \frac{1}{2} e^{i \tilde{x}_{L} p^+ y_0^-} e^{i (x_{D1}-x_M) p^+ \delta y_1^-} e^{-i \tilde{x}_L p^+ (Y_1^-+\delta y_1^-/2)}\approx - \frac{1}{2} e^{-i \tilde{x}_L p^+ Y_1^-},
\nonumber\\
S_{(\ref{fig6})}&=&S_{(\ref{fig6},a)}+S_{(\ref{fig6},b)}\approx - \cos(\tilde{x}_L p^+ Y_1^-),
\end{eqnarray}
in which
\begin{eqnarray}
x_{D1}=\frac{{p}_{1\perp}^2+M^2}{2 p^+ q^- (1+x_1^-)}.
\end{eqnarray}
The hard part of the matrix element for Fig.~\ref{fig6} reads:
\begin{eqnarray}
{T}_{(\ref{fig6},a)} ={T}_{(\ref{fig6},b)} &=&{2yP(y)}\frac{l_\perp^2+\frac{y^4}{1+(1-y)^2}M^2}{(l_\perp^2+y^2 M^2)^2}.
\end{eqnarray}

The phase factor for Fig.~\ref{fig7} reads:
\begin{eqnarray}
S_{(\ref{fig7},a)}&=&- \frac{1}{2}\left(e^{i \tilde{x}_L p^+ y_2^-}e^{-i (x_{D1}'-x_M) p^+ (y_2'^--y_1'^-)} - e^{i \tilde{x}_L p^+ y_1'^-}e^{-i[x_{C1}'-y\tilde{x}_L-(1+y)x_M]p^+(y_2'^- -y_1'^-)}\right)
\nonumber\\
&\approx&- \frac{1}{2}\left(e^{i \tilde{x}_L p^+ (Y_1^-+\delta y_1^-/2)}e^{-i (x_{D1}-x_M) p^+ \delta y_1^-} - e^{i \tilde{x}_L p^+ (Y_1^-+\delta y_1^-/2)}e^{-i[x_{C1}-y\tilde{x}_L-(1+y)x_M]p^+\delta y_1^-}\right)\approx\;0,
\nonumber\\
S_{(\ref{fig7},b)}&=&- \frac{1}{2}e^{- i \tilde{x}_L p^+ y_0^-}\left(e^{- i \tilde{x}_L p^+ y_2^-}e^{i (x_{D1}-x_M) p^+ (y_2^--y_1^-)} - e^{- i \tilde{x}_L p^+ y_1^-}e^{i[x_{C1}-y\tilde{x}_L-(1+y)x_M]p^+(y_2^- -y_1^-)}\right)
\nonumber\\
&\approx&- \frac{1}{2}e^{i \tilde{x}_L p^+ y_0^-}\left(e^{- i \tilde{x}_L p^+ (Y_1^-+\delta y_1^-/2)}e^{i (x_{D1}-x_M) p^+ \delta y_1^-} - e^{- i \tilde{x}_L p^+ (Y_1^-+\delta y_1^-/2)}e^{i[x_{C1}-y\tilde{x}_L-(1+y)x_M]p^+\delta y_1^-}\right)\approx\;0,
\nonumber\\
S_{(\ref{fig7})}&=&S_{(\ref{fig7},a)}\,\,+\,\,S_{(\ref{fig7},b)}\,\approx\,0.
\end{eqnarray}
The hard part of the matrix element for Fig.~\ref{fig7} reads:
\begin{eqnarray}
{T}_{(\ref{fig7},a)}={T}_{(\ref{fig7},b)}= {2yP(y)} \frac{1+(1-y)\left(1-\frac{y}{1+x_1^-}\right)}{1+(1-y)^2}
\frac{\mathbf{l}_{\perp} \cdot \left(\mathbf{l}_{\perp} - \frac{y}{1+ x_1^-} \mathbf{p}_{1\perp}\right)+\frac{y^2\left(\frac{y}{1+x_1^-}\right)^2 M^2}{1+(1-y)\left(1-\frac{y}{1+x_1^-}\right)}}{(l_\perp^2 + y^2 M^2)
\left[\left(\mathbf{l}_{\perp} - \frac{y}{1+ x_1^-} \mathbf{p}_{1\perp}\right)^2+\left(\frac{y}{1+ x_1^-}\right)^2 M^2\right]}.
\end{eqnarray}

%\end{widetext}

%%%%%%%%%%%%%%%%%%%%%%%%%%%%%%%%%%%%%%%%%%%%%%%%%%%%%%%%%%%%%%%%%%%%%%%%%%%%%%%
%%%%%%%%%%%%%%%%%%%%%%%%%%%%%%%%%%%%%%%%

%%%%%%%%%%%%%%%%%%%%%%%%%%%%%%%%%%%%%%%%%%%%%%%%%%%%%%%%%%%%%%%%%%%%%%%%%%%%%%%%
%%%%%%%%%%%%%%%%%%%%%%%%%%%%%%%%%%%%%%%%%
\bibliographystyle{plain}
\bibliographystyle{h-physrev5}
\bibliography{refs_GYQ}
%%%%%%%%%%%%%%%%%%%%%%%%%%%%%%%%%%%%%%%%%%%%%%%%%%%%%%%%%%%%%%%%%%%%%%%%%%%%%%%%
%%%%%%%%%%%%%%%%%%%%%%%%%%%%%%%%%%%%%%%%
\end{document}